\documentclass[%
 reprint,
 amsmath,amssymb,
 aps,
 prl,
 longbibliography,
 lengthcheck,%
]{revtex4-1}

\usepackage{graphicx}
\usepackage{dcolumn}
\usepackage{bm}
\usepackage{hyperref}

\begin{document}

\preprint{APS/123-QED}

\title{A quantum hydrodynamics approach to the formation of new types of waves in polarized
 two-dimension systems of charged and neutral particles}

\author{P. A. Andreev}
\email{andrap@yandex.ru}
 \affiliation{Department of General Physics, Physics Faculty, Moscow State
University, Moscow, Russian Federation.}
\author{L. S. Kuzmenkov}%
 \email{lsk@phys.msu.ru}
\author{M. I. Trukhanova}
  \email{mar-tiv@yandex.ru}
\affiliation{%
 Department of Theoretical Physics, Physics Faculty, Moscow State
University, Moscow, Russian Federation.
}%

\date{\today}

\begin{abstract}
In this paper we explicate a method of quantum hydrodynamics (QHD)
for the study of the quantum evolution of a system of polarized
particles. Though we focused primarily on the two-dimension
physical systems, the method is valid for three-dimension and
one-dimension systems too. The presented method is based upon the
Schr\"{o}dinger equation. Fundamental QHD equations for charged
and neutral particles were derived from the many-particle
microscopic Schr\"{o}dinger equation. The fact that particles
possess the electric dipole moment (EDM) was taken into account.
The explicated QHD approach was used to study dispersion
characteristics of various physical systems. We analyzed
dispersion of waves in a two-dimension (2D) ion and hole gas
placed into an external electric field which is orthogonal to the
gas plane. Elementary excitations in a system of neutral polarized
particles were studied for 1D, 2D and 3D cases. The polarization
dynamics in systems of both neutral and charged particles is shown
to cause formation of a new type of waves as well as changes in
the dispersion characteristics of already known waves. We also
analyzed wave dispersion in 2D exciton systems, in 2D electron-ion
plasma and 2D electron-hole plasma. Generation of waves in 3D
system neutral particles with EDM by means of the beam of
electrons and neutral polarized particles is investigated.

\end{abstract}

\pacs{73.20.Mf, 47.10.Fg, 52.35.Fp}
\maketitle


\section{\label{sec:level1}I. Introduction}
A two-dimension electron gas (2DEG) is in the focus  of many
studies. Sometimes it is considered to be an element of the
spin-field effect transistor  ~\cite{Zutic RMP 04}. Ions and holes are
distinct from electrons in that they can polarize in an external
electric field or arrange in the line of the external field in the
case they are rigid dipoles. EDM of ions and holes may cause changes in the
dispersion of known eigenwaves as well as the onset of waves of
novel types.

Dispersion characteristics of 2DEG have been analyzed in works of
~\cite{Oji PRB 86,Batke PRB 86,Xu PRB 06}. Magnetoplasma  waves in
2DEG have been studied in  ~\cite{Oji PRB 86,Batke PRB 86}.
Excitation of charge and spin density in 2DEG has been addressed
in ~\cite{Xu PRB 06} where spin-orbital interaction has been taken
into account and the 2DEG has been supposed to locate in an
external magnetic field. Not only wave processes attract
researchers' interest, but 2D magneto-transport also does. Studies
of the effect of spin-orbital interaction on the 2D
magneto-transport are also important. An equation set has been
presented in ~\cite{Burkov PRB 04} that describes charge and spin
diffusion in 2DEG with account of spin-orbital interaction
~\cite{Miller PRL 03}. 2D hole gas along with 2DEG is used for
construction of transistors, for example Atomic Layer Doping-Field
Effect Transistor (ALD-FET) ~\cite{ALD-FET}. Collective modes in
2D bilayer graphene are considered in article ~\cite{Cote PRB 10}.

In recent years attention is paid to the effect of  the intrinsic
magnetic moment (IMM) on the characteristics of charged particle
systems. The propagation of perturbation of IMM and EDM does not require much
energy as it occurs without mass transfer. Both processes may be
used in the information transfer. In biological systems, for
example, polarization processes, i.e. EDM propagation, are the
predominant way of signal transfer.

There are modern studies which are focused primarily on the EDM  dynamics ~\cite{Andreev DSS 10}, ~\cite{Qiuzi Li PRB 11}. They
deal with quasi-2D and multilayered systems of both charged and
neutral polarized particles. There are articles where take into account of the EDM influence
on the Bose-Einstein condensate (BEC) ~\cite{Ticknor PRL 11}, ~\cite{Andreev arxiv 11 2}. In a particular work a
contribution of polarization to the dispersion of Bogoliubov's
mode has been analyzed.

Influence of EDM on dynamic of charged particles is not large, analogously with the influence of IMM. But, evolution of IMM lead to existence of new physical effects which described below. In this way we can expect the existence new physical process which are the consequence of EDM dynamics.
IMM dynamics causes changes in the wave dispersion in magnetized
plasma ~\cite{Marklund PRL07,Andreev VestnMSU 2007,Andreev AtPhys
08} as well as the existence of novel branches of dispersion in
systems of such kind ~\cite{Andreev VestnMSU 2007}. Interaction
between IMM of neutron beam and magnetized plasma leads to the
generation of waves in plasma~\cite{Andreev arxiv 11 2}, ~\cite{Andreev AtPhys
08}. This effect is due to spin-spin,
spin-current ~\cite{Andreev AtPhys 08} and spin-orbital
interactions ~\cite{Andreev PIERS 2011} between IMM of the beam
and IMM and currents in plasma. A lot of works is also devoted to
the dynamics of IMM in BEC.

Classical methods were used previously to create a description of
the collective dynamics of charged particles that takes into
account the EDM that arises in the medium as a consequence of
charged particles' movement ~\cite{Drofa TMP 96}.

A quantum mechanics description for systems of $N$ interacting
particles is based upon the many-particle Schr\"{o}dinger equation
(MPSE) that specifies a wave function in a 3N-dimensional
configuration space. As wave processes, processes of information
transfer, diffusion and other transport processes occur in the
three-dimensional physical space, a need arises to turn to a
mathematical method of physically observable values which are
determined in a 3D physical space. To do so we should derive
equations those determine dynamics of functions of three
variables, starting from MPSE. This problem has been solved with
the creation of a method of many-particle quantum hydrodynamics
(MPQHD). Therefore, the reason for development such method is analogous with the motivation of the density functional
theory ~\cite{Kohn RMP 99}.
In our work we propose a further development of the MPQHD
method. Here we consider the EDM dynamics in systems of charged
and neutral particles. The MPQHD method has been applied to
systems of charged ~\cite{MaksimovTMP 1999,MaksimovTMP
2001,MaksimovTMP 2001(2)}, ~\cite{Andreev Izv.Vuzov. 07} and neutral ~\cite{Andreev PRA08}
particles before. Many researchers ~\cite{Haas PP05,Marklund
PRL07,Brodin NJP07}, ~\cite{Asenjo PP 11} e.g. Madelung ~\cite{Madelung 1926} and
Takabayashi ~\cite{Takabayasi} consider derivation of QHD equations from a
Schr\"{o}dinger equation for a single particle in an external
field. Marklund et. al. ~\cite{Marklund PRL07} suggests the
method of generalization of one-particle QHD equation for
description of many-particle systems. In article ~\cite{Asenjo PP
11} method proposed in ~\cite{Marklund PRL07} is used for
derivation of QHD equations for system of relativistic particles.
Both classical and quantum dynamics of a plenty of interacting
particles in a configuration space are discussed in
~\cite{Albareda PRB 09}.

Bibliography contains many examples of QHD application  for the
analysis of various processes and phenomena. Ion-acoustic waves in
dusty plasma ~\cite{S. A. Khan PP 08,S. A. Khan PP 07,S Ali NJoP
08} and the distribution of non-linear electrostatic solitary
excitations in it ~\cite{M. Akbari PP 10} have been analyzed. QHD
method is also used to study wave processes in
electron-positron-ion plasma ~\cite{S. Ali PP 07,S. A. Khan CPL
09,Haijun Ren JP. A 08}. A system of QHD equations can be applied
to the analysis of instability of quantum plasma ~\cite{F. Haas
PRE 08,L. A. Rios PP 08,A. P. Misra PP 08,Haijun Ren PP 09,Mikhail
Modestov PP 09}, and in particular of the modulation instability
of electron cyclotron waves ~\cite{L. A. Rios PP 08} and
magnetoacoustic waves ~\cite{A. P. Misra PP 08}. Dispersion
characteristics of magnetoacoustic waves have been analyzed in
this way too ~\cite{W. Masood PP 09,M. Marklund PRE 07,A. Mushtaq
PP 09}. The dispersion of electrostatic waves with frequencies
below the electron plasma frequency has also been studied using
QHD ~\cite{H. Tercas PP 08}. The dispersion relation for ion
acoustic waves and the absorption coefficient for Landau damping
are obtain  in paper ~\cite{L. A. Rios PP 08} and the quantum
electrodynamical short wavelength correction on plasma wave
propagation for a nonrelativistic quantum plasma is investigated
in article ~\cite{J. Lundin PP 07}.

Some of the results that have been obtained using  QHD equations
derived from a single-particle Schr\"{o}dinger equation are
presented in the review ~\cite{P.K. Shukla UFN 10}.

The MPQHD approach to systems of neutral particles,  e.g. of
ultracold boson-fermion mixtures, has been developed in
~\cite{Andreev PRA08} and applied later to describe linear and
non-linear characteristics of BEC and of boson-fermion mixtures
~\cite{Andreev PRA08,Andreev Izv.Vuzov. 09 1,Andreev Izv.Vuzov.
10,Andreev arxiv 11 1}.

In this work we extend MPQHD approach to  a system of EDM-having
particles. The method we developed here may be an effective tool
for the investigation of static and dynamic behavior and transport
characteristics of  in graphene ~\cite{Das Sarma RMP 11} and
nanofluidics ~\cite{Schoch RMP 08}, the last one deals with the
dynamics of ions in water, a substance with large EDM of
molecules.

In this article, using the QHD  approach we calculate the
dispersion of polarization waves and spatial charge waves. We show
the EDM dynamics to cause novel branches of dispersion in quasi-2D
plasma-like media and in systems of neutral particles. We also
accomplish analytical calculations of dispersion characteristics
for the waves we discovered.

Electrically polarized particle system can interact with the beam
of charged and polarized particles by means charge-dipole and
dipole-dipole interaction. Such interaction could lead to transfer
energy from beam to medium and, consequently, to generation of
waves. In plasma physics the effect of generation of waves by
means electron ~\cite{Bret PRE 04}, ~\cite{Miller BOOK} or
magnetized neutron ~\cite{Andreev PIERS 2011}  beam is well-known.
In presented article we consider similar effects in system neutral
polarized particles.

Our paper is organized as follows. In Sect. 2 we present the
derivation of the momentum balance equation for EDM-having charged
particles from MPSE in a self-consistent field approximation. An
explicit form of the quantum part of the pressure tensor is also
presented. In Sect. 3 we obtain equations of polarization
evolution and polarization current. The self-consistent field
approximation is used.   In Sect. 4 a calculation is accomplished
of eigenwaves in a 2D system of EDM-having charged particles
located in uniform external electric field. New dispersion branch
of $\omega(\textbf{k})$ is shown to exist and the contribution of
polarization into the dispersion of 2D Langmuire wave is
estimated. In Sect. 5 we show the existence of a polarization wave
along with acoustic wave in a system of neutral polarized
particles for 1D, 2D and 3D cases.  In Sect. 6 dispersion
characteristics of a two-sort 2D system of charged particles are
discussed. An assumption is made that particles of one of the
sorts bear EDM. We show that polarization dynamics here leads to
the existence of a new dispersion branch. Analytical relation for
$\omega(\textbf{k})$ is constructed. Deep analysis of quantum
magnetoacoustic waves is performed. In Sect. 7 we apply the
results presented in sections 4-6 to the analysis of excitations
in 2D electron plasma and in a system of excitons.  In Sect. 8 we
show that there is instability at interaction of beam of polarized
particles with the polarized medium. The increment of
instabilities is calculated.  In Sect. 9 we obtain the increment
of instabilities which arise at electron beam propagation through
system of neutral particles with the EDM. In sect. 10 brief
summary of obtained results are presented.

\section{\label{sec:level1} II. Construction of fundamental equations and the model accepted}

In this section we derive the MPQHD equations from MPSE. Here we
present the key steps of getting the MPQHD equations. We receive
the equations for the system of charged particles with EDM. Obtaining
equations could be used for neutral particles with the EDM as well.
Method of MPQHG allow to present dynamic of system of interacting
quantum particles in terms of functions defined in 3D physical space.
It is important at investigation of wave process, which take place in 3D physical space.

Starting from MPSE
$$-\imath\hbar\partial_{t}\psi=\hat{H}\psi$$
with the Hamiltonian
$$\hat{H}=\sum_{i}\Biggl(\frac{1}{2m_{i}}\textbf{D}_{i}^{2}+e_{i}\varphi_{i,ext}-d_{i}^{\alpha}E_{i,ext}^{\alpha}\Biggr)$$
\begin{equation}\label{di Hamiltonian}+\sum_{i,j\neq i}\Biggl(\frac{1}{2}e_{i}e_{j}G_{ij}+e_{i}d_{j}^{\alpha}C_{ij}^{\alpha}-\frac{1}{2}d_{i}^{\alpha}d_{j}^{\beta}G_{ij}^{\alpha\beta}\Biggr),\end{equation}
we construct a system of QHD equations for charged particles that
have EDM $d_{i}^{\alpha}$. Equations constructed in this way are
also valid for neutral polarized particles. The following
designations are used in the Hamiltonian (\ref{di Hamiltonian}):
$D_{i}^{\alpha}=-\imath\hbar\partial_{i}^{\alpha}-e_{i}A_{i,ext}^{\alpha}/c$,
$\varphi_{i,ext}$, $A_{i,ext}^{\alpha}$ - the potentials of
external electromagnetic field, $E_{i,ext}^{\alpha}$-, is the
electric field, quantities $e_{i}$, $m_{i}$-are the charge and
mass of particles, $\hbar$-is the Planck constant, and
$G_{ij}=1/r_{ij}$,
$C_{ij}^{\alpha}=-\partial_{i}^{\alpha}1/r_{ij}$,
$G_{ij}^{\alpha\beta}=\partial_{i}^{\alpha}\partial_{i}^{\beta}1/r_{ij}$
- the Green functions of the Coulomb, charge-dipole, dipole-dipole
interactions, respectively.

The first step in the construction of QHD apparatus is to determine the concentration of particles in the neighborhood of $\textbf{r}$ in a physical space. If we define the concentration of particles as quantum average of the concentration operator in the coordinate representation $\hat{n}=\sum_{i}\delta(\textbf{r}-\textbf{r}_{i})$ we obtain:
\begin{equation}\label{di def density}n(\textbf{r},t)=\int dR\sum_{i}\delta(\textbf{r}-\textbf{r}_{i})\psi^{*}(R,t)\psi(R,t)\end{equation}
where $dR=\prod_{p=1}^{N}d\textbf{r}_{p}$.

Differentiation of $n(\textbf{r},t)$ with respect to time and applying of the Schr\"{o}dinger equation with Hamiltonian (\ref{di Hamiltonian}) leads to
continuity equation
\begin{equation}\label{di continuity equation}\partial_{t}n(\textbf{r},t)+\nabla\textbf{j}(\textbf{r},t)=0\end{equation}
where the current density takes a form of
$$j^{\alpha}(\textbf{r},t)=\int dR\sum_{i}\delta(\textbf{r}-\textbf{r}_{i})\frac{1}{2m_{i}}\biggl(\psi^{*}(R,t)(D_{i}^{\alpha}\psi(R,t))$$
\begin{equation}\label{di def of current of density}+(D_{i}^{\alpha}\psi(R,t))^{*}\psi(R,t)\biggr),\end{equation}

The velocity of i-th particle $\textbf{v}_{i}(R,t)$ is determined by equation
\begin{equation}\label{di vel of i part}\textbf{v}_{i}(R,t)=\frac{1}{m_{i}}\nabla_{i}S(R,t)-\frac{e_{i}}{m_{i}c}\textbf{A}_{i, ext}
.\end{equation}
The quantity $\textbf{v}_{i}(R,t)$ describe the current of probability connected with the motion of $i$-th particle, in general case $\textbf{v}_{i}(R,t)$ depend on coordinate of all particles of the system $R$, where $R$ is the totality of 3N coordinate of N particles of the system $R=(\textbf{r}_{1}, ..., \textbf{r}_{N}, t)$.

The $S(R,t)$ value in the formula (\ref{di vel of i part}) represents the phase of the wave function
$$\psi(R,t)=a(R,t) exp(\frac{\imath S(R,t)}{\hbar}).$$

Velocity field $\textbf{v}(\textbf{r},t)$ is the velocity of the
local centre of mass and determined by equation:
\begin{equation}\label{di current to velocity}\textbf{j}(\textbf{r},t)=n(\textbf{r},t)\textbf{v}(\textbf{r},t).\end{equation}

This means that $\textbf{u}_{i}(\textbf{r},R,t)=\textbf{v}_{i}(R,t)-\textbf{v}(\textbf{r},t)$ is a quantum equivalent of the thermal speed.

A momentum balance equation can be derived by differentiating current density (\ref{di def of current of density}) with respect to time:
 \begin{equation} \label{di bal of imp gen}\partial_{t}j^{\alpha}(\textbf{r},t)+\frac{1}{m}\partial^{\beta}\Pi^{\alpha\beta}(\textbf{r},t)=F^{\alpha}(\textbf{r},t),\end{equation}
where $F^{\alpha}(\textbf{r},t)$ is a force field and
$$\Pi^{\alpha\beta}(\textbf{r},t)=\int
 dR\sum_{i}\delta(\textbf{r}-\textbf{r}_{i})\times$$
 $$\times\frac{1}{4m_{i}}\biggl(\psi^{+}(R,t)(\hat{D}^{\alpha}_{i}\hat{D}^{\beta}_{i}\psi)(R,t)+$$
 \begin{equation} \label{di Pi}+(\hat{D}^{\alpha}_{i}\psi)^{+}(R,t)(\hat{D}^{\beta}_{i}\psi)(R,t)+c.c.\biggr) \end{equation}
represents the momentum current density tensor.

Let's now perform explicit separation of particles' thermal movement with velocities $\textbf{u}_{i}(\textbf{r},R,t)$ and the collective movement of particles with velocity $\textbf{v}(\textbf{r},t)$ in equations of continuity (\ref{di continuity equation}) and of the momentum balance (\ref{di bal of imp gen}). We can see now that the tensor $\Pi^{\alpha\beta}(\textbf{r},t)$ takes the form
$$\Pi^{\alpha\beta}(\textbf{r},t)=mn(\textbf{r},t)v^{\alpha}(\textbf{r},t)v^{\beta}(\textbf{r},t)$$
$$+p^{\alpha\beta}(\textbf{r},t)+T^{\alpha\beta}(\textbf{r},t).$$

In this formula
\begin{equation}\label{di pressure} p^{\alpha\beta}(\textbf{r},t)=\int dR\sum_{i=1}^{N}\delta(\textbf{r}-\textbf{r}_{i})a^{2}(R,t)m_{i}u^{\alpha}_{i}u^{\beta}_{i} \end{equation}
is the tensor of kinetic pressure. This tensor tends to zero by
letting $\textbf{u}_{i}\rightarrow 0$.

The tensor
$$T^{\alpha\beta}(\textbf{r},t)=-\frac{\hbar^{2}}{2m}\times$$
\begin{equation}\label{di Bom1} \times\int dR\sum_{i=1}^{N}\delta(\textbf{r}-\textbf{r}_{i})a^{2}(R,t)\frac{\partial^{2}\ln a(R,t)}{\partial x_{\alpha i}\partial x_{\beta i}}
\end{equation}
is proportional to $\hbar^{2}$ and has a purely
quantum origin. For the large system of noninteracting particles,
this tensor is

$$T^{\alpha\beta}(\textbf{r},t)=-\frac{\hbar^{2}}{4m}\partial^{\alpha}\partial^{\beta}n(\textbf{r},t)$$
\begin{equation}\label{di Bom2}+\frac{\hbar^{2}}{4m}\frac{1}{n(\textbf{r},t)}(\partial^{\alpha}n(\textbf{r},t))(\partial^{\beta}n(\textbf{r},t))
.\end{equation} This term is named Bohm quantum potential.

As the particles of the system under consideration interact via long-range forces the approximation of a self-consistent field is sufficient to analyze collective processes. With the use of this approximation two-particle functions in the momentum balance equation can be split into a product of single-particle functions. Taken in the approximation of self-consistent field, the set of QHD equation, continuity equation and momentum balance equation has a form:

\begin{equation}\label{di cont eq}
\partial_{t}n(\textbf{r},t)+\nabla(n(\textbf{r},t)\textbf{v}(\textbf{r},t))=0 ,\end{equation}
$$mn(\textbf{r},t)(\partial_{t}+\textbf{v}\nabla)v^{\alpha}(\textbf{r},t)+\partial_{\beta}(p^{\alpha\beta}(\textbf{r},t)+T^{\alpha\beta}(\textbf{r},t))
$$
$$=en(\textbf{r},t)E_{ext}^{\alpha}(\textbf{r},t)+
P^{\beta}(\textbf{r},t)\partial^{\alpha}E_{ext}^{\beta}(\textbf{r},t)$$
$$+en(\textbf{r},t)\varepsilon^{\alpha\beta\gamma}v^{\beta}(\textbf{r},t)B_{ext}^{\gamma}(\textbf{r},t)$$
$$-e^{2}n(\textbf{r},t)\partial^{\alpha}\int
d\textbf{r}'G(\textbf{r},\textbf{r}')n(\textbf{r}',t)$$
$$-en(\textbf{r},t)\partial^{\alpha}\partial^{\beta}\int d\textbf{r}'G(\textbf{r},\textbf{r}')P^{\beta}(\textbf{r}',t)$$
$$+eP^{\beta}(\textbf{r},t)\partial^{\alpha}\partial^{\beta}\int
d\textbf{r}'G(\textbf{r},\textbf{r}')n(\textbf{r}',t)$$
\begin{equation}\label{di bal imp eq}+P^{\beta}(\textbf{r},t)\partial^{\alpha}\int d\textbf{r}'G^{\beta\gamma}(\textbf{r},\textbf{r}')P^{\gamma}(\textbf{r}',t).\end{equation}

Let's discuss the physical significance of terms on the right side of (\ref{di bal imp eq}). The first three terms describe the interaction with external electromagnetic field. The first term represents the effect of external electric field on the charge density. The second term is the effect of non-uniform external electric field on the polarization density. The form of this term is similar to that of the force field which affects the magnetic moment in a magnetic field ~\cite{MaksimovTMP 2001}, ~\cite{Takabayasi}. It should be noted that the form of this term is distinct from the expression that describes the force affecting a single dipole. The third term of the equation (\ref{di bal imp eq}) is the Lorentz force. Other terms in (\ref{di bal imp eq}) describe a force field that represents interactions between particles, namely the Coulomb interaction of charges, the effect of dipoles on charges, charges on dipoles and dipole-dipole interactions.

Note, that for a 3D system of particles the momentum balance equation (\ref{di bal imp eq}) may be written down in terms of electrical intensity of the field that is created by charges and dipole moments of the particle system:
$$mn(\textbf{r},t)(\partial_{t}+\textbf{v}\nabla)v^{\alpha}(\textbf{r},t)+\partial_{\beta}(p^{\alpha\beta}(\textbf{r},t)+T^{\alpha\beta}(\textbf{r},t))
$$
$$=en(\textbf{r},t)E^{\alpha}(\textbf{r},t)+
P^{\beta}(\textbf{r},t)\partial^{\alpha}E^{\beta}(\textbf{r},t)$$
\begin{equation}\label{di bal imp eq short}+en(\textbf{r},t)\varepsilon^{\alpha\beta\gamma}v^{\beta}(\textbf{r},t)B_{ext}^{\gamma}(\textbf{r},t),\end{equation}
where:
$E^{\alpha}(\textbf{r},t)=E_{ext}^{\alpha}(\textbf{r},t)+E_{int}^{\alpha}(\textbf{r},t)$
and
$E_{int}^{\alpha}(\textbf{r},t)=E_{q}^{\alpha}(\textbf{r},t)+E_{d}^{\alpha}(\textbf{r},t)$.
These variables meet equations $div\textbf{E}_{q}(\textbf{r},t)=4\pi\rho$ and $div\textbf{E}_{d}(\textbf{r},t)=-4\pi
div\textbf{P}(\textbf{r},t)$ where $\rho=\sum_{a}e_{a}n_{a}(\textbf{r},t)$.
This leads to a field equation
\begin{equation}\label{di field eq}div\textbf{E}_{int}(\textbf{r},t)=4\pi\rho-4\pi div\textbf{P}.\end{equation}

The method we develop in this work is valid both for bosons and fermions. The type of statistics that particles are subject to affects the calculation of many-particle functions (correlations) that evolve in the momentum balance equation (\ref{di bal imp eq}) and are neglected in the self-consistent field approximation. A method for the calculation of correlations in QHD equations has been developed in works ~\cite{MaksimovTMP 1999,MaksimovTMP 2001,MaksimovTMP 2001(2)}.

Polarization evolves in the momentum balance equation (\ref{di bal imp eq}) or (\ref{di bal imp eq short}):
\begin{equation}\label{di def polarization}P^{\alpha}(\textbf{r},t)=\int dR\sum_{i}\delta(\textbf{r}-\textbf{r}_{i})\psi^{*}(R,t)\hat{d}_{i}^{\alpha}\psi(R,t).\end{equation}

The EDM operator $d_{i}^{\alpha}$ there affects coordinates of the i-th particle. The polarization thus affects the evolution of concentration and of the velocity field. To calculate $P^{\alpha}(\textbf{r},t)$ may be interesting itself. It is, therefore necessary to derive an equation that describes the evolution of polarization.

\section{\label{sec:level1} III. Equations for the evolution of polarization}

To close the QHD equations set (\ref{di cont eq}), (\ref{di bal imp eq}) we derive equation for the polarization evolution. If we differentiate the definition for polarization (\ref{di def polarization}) with respect to time and apply the Schr\"{o}dinger equation, the required equation for the polarization evolution can be obtained:
\begin{equation}\label{di eq polarization}\partial_{t}P^{\alpha}(\textbf{r},t)+\partial^{\beta}R^{\alpha\beta}(\textbf{r},t)=0.\end{equation}

A polarization current density
$$R^{\alpha\beta}(\textbf{r},t)=\int dR\sum_{i}\delta(\textbf{r}-\textbf{r}_{i})\frac{d_{i}^{\alpha}}{2m_{i}}\times$$
\begin{equation}\label{di def of current of polarization}\times\biggl(\psi^{*}(R,t)(D_{i}^{\beta}\psi(R,t))+(D_{i}^{\beta}\psi(R,t))^{*}\psi(R,t)\biggr)\end{equation}
 occurs in this equation.

We have two ways to close the QHD equations set. The first one is
to express $R^{\alpha\beta}(\textbf{r},t)$ in terms of
$n(\textbf{r},t)$, $v^{\alpha}(\textbf{r},t)$ and
$P^{\alpha}(\textbf{r},t)$ using additional assumptions or
experimental data. The other way is to derive the equation for
evolution $R^{\alpha\beta}(\textbf{r},t)$  in the same fashion it
was accomplished previously for other material fields, for example
quantities considered above $n(\textbf{r},t)$,
$v^{\alpha}(\textbf{r},t)$ and $P^{\alpha}(\textbf{r},t)$. Now the
evolution equation $R^{\alpha\beta}(\textbf{r},t)$  occurs in the
form of
\begin{widetext}
$$\partial_{t}R^{\alpha\beta}(\textbf{r},t)+\frac{1}{m}\partial^{\gamma}R^{\alpha\beta\gamma}(\textbf{r},t)=\frac{e}{m}P^{\alpha}(\textbf{r},t)E^{\beta}_{ext}(\textbf{r},t)$$
$$+\frac{e}{mc}\varepsilon^{\beta\gamma\delta}R^{\alpha\gamma}(\textbf{r},t)B^{\delta}_{ext}(\textbf{r},t)+\frac{1}{m}(\partial^{\beta}E^{\gamma}_{ext}(\textbf{r},t))\int dR\sum_{p}\delta(\textbf{r}-\textbf{r}_{p})d_{p}^{\alpha}d_{p}^{\gamma}\psi^{*}(R,t)\psi(R,t)$$
$$-\frac{e^{2}}{m}\int d\textbf{r}'(\partial^{\beta}G(\textbf{r},\textbf{r}'))\int dR\sum_{p,n\neq p}\delta(\textbf{r}-\textbf{r}_{p})\delta(\textbf{r}-\textbf{r}_{n})d_{p}^{\alpha}\psi^{*}(R,t)\psi(R,t)$$
$$-\frac{e}{m}\int d\textbf{r}'(\partial^{\beta}G^{\gamma}(\textbf{r}',\textbf{r}))\int dR\sum_{p,n\neq p}\delta(\textbf{r}-\textbf{r}_{p})\delta(\textbf{r}-\textbf{r}_{n})d_{p}^{\alpha}d_{p}^{\gamma}\psi^{*}(R,t)\psi(R,t)$$
$$+\frac{e}{m}\int d\textbf{r}'(\partial^{\beta}G^{\gamma}(\textbf{r},\textbf{r}'))\int dR\sum_{p,n\neq p}\delta(\textbf{r}-\textbf{r}_{p})\delta(\textbf{r}-\textbf{r}_{n})d_{p}^{\alpha}d_{n}^{\gamma}\psi^{*}(R,t)\psi(R,t)$$
\begin{equation}\label{di eq for pol current 2}+\frac{1}{m}\int d\textbf{r}'(\partial^{\beta}G^{\gamma\delta}(\textbf{r},\textbf{r}'))\int dR\sum_{p,n\neq p}\delta(\textbf{r}-\textbf{r}_{p})\delta(\textbf{r}-\textbf{r}_{n})d_{p}^{\alpha}d_{p}^{\gamma}d_{n}^{\delta}\psi^{*}(R,t)\psi(R,t).\end{equation}
\end{widetext}

Now we consider the physical meaning of the terms in right side of this equation.
The first three terms describe the interaction of particles
with external electromagnetic field. The first term
represents the effect of external electric field on the charge
density. The second term is the effect of non-uniform external
electric field on the polarization density. The form of
this term is similar to that of the force field which affects
the magnetic moment in a magnetic field. The third term of
the equation (\ref{di eq for pol current 2}) is an analog of the Lorentz
force. Other terms in (\ref{di eq for pol current 2}) describe a tensor field that represents
interactions between particles, namely the Coulomb
interaction of charges, the effect of dipoles on charges,
charges on dipoles and dipole-dipole interactions.

We used the denotation in (\ref{di eq for pol current 2}) for $R^{\alpha\beta\gamma}(\textbf{r},t)$:
$$R^{\alpha\beta\gamma}(\textbf{r},t)$$
$$=\int
 dR\sum_{i}\delta(\textbf{r}-\textbf{r}_{i})\frac{d_{i}^{\alpha}}{4m_{i}}\biggl(\psi^{+}(R,t)(\hat{D}^{\beta}_{i}\hat{D}^{\gamma}_{i}\psi)(R,t)+$$
\begin{equation} \label{di Pi for polarization}
+(\hat{D}^{\beta}_{i}\psi)^{+}(R,t)(\hat{D}^{\gamma}_{i}\psi)(R,t)+c.c.\biggr).
\end{equation}

Equation (\ref{di eq for pol current 2}) show the evolution of
polarization current $R^{\alpha\beta}(\textbf{r},t)$  in consequence
of interaction of particles with external and internal force field.
Equation (\ref{di eq for pol current 2}) contain two-particle
correlation function. The two-particle function found in the
four last term of equation (\ref{di eq for pol current 2}). In
articles ~\cite{MaksimovTMP 1999}, ~\cite{MaksimovTMP 2001(2)}
the method of calculation of correlation is developed. Direct calculation of many-particle function using methods developed in ~\cite{MaksimovTMP 1999}, ~\cite{MaksimovTMP 2001(2)} show that statistics affect on correlations only and have no influence on terms arisen in self-consistent field approximation. Below, in this article we bound our self by self-consistent field approximation.

We can see that the equation (\ref{di eq for pol current 2})
contains information about the effect of interaction on the
transformation of $R^{\alpha\beta}(\textbf{r},t)$ and, as a
consequence, on the polarization evolution
$P^{\alpha}(\textbf{r},t)$.

In the approximation of the self-consistent field, the
equation (\ref{di eq for pol current 2}) takes the form
$$\partial_{t}R^{\alpha\beta}(\textbf{r},t)+\frac{1}{m}\partial^{\gamma}R^{\alpha\beta\gamma}(\textbf{r},t)=\frac{e}{m}P^{\alpha}(\textbf{r},t)E_{ext}^{\beta}(\textbf{r},t)$$
$$+\frac{e}{mc}\varepsilon^{\beta\gamma\delta}R^{\alpha\gamma}(\textbf{r},t)B_{ext}^{\delta}(\textbf{r},t)+\frac{1}{m}D^{\alpha\gamma}(\textbf{r},t)\partial^{\beta}E^{\gamma}_{ext}(\textbf{r},t)$$
$$-\frac{e^{2}}{m}P^{\alpha}(\textbf{r},t)\partial^{\beta}\int
d\textbf{r}'G(\textbf{r},\textbf{r}')n(\textbf{r}',t)$$
$$-\frac{e}{m}D^{\alpha\gamma}(\textbf{r},t)\partial^{\beta}\int d\textbf{r}'G^{\gamma}(\textbf{r}',\textbf{r})n(\textbf{r}',t)$$
$$+\frac{e}{m}P^{\alpha}(\textbf{r},t)\partial^{\beta}\int d\textbf{r}'G^{\gamma}(\textbf{r},\textbf{r}')P^{\gamma}(\textbf{r}',t)$$
\begin{equation}\label{di eq for pol current selfconsist appr}+\frac{1}{m}D^{\alpha\gamma}(\textbf{r},t)\partial^{\beta}\int d\textbf{r}'G^{\gamma\delta}(\textbf{r},\textbf{r}')P^{\delta}(\textbf{r}',t).\end{equation}
The physical meaning of terms on the right side of this equation
is similar to that of terms in the equation (\ref{di eq for pol current 2}).

We used the denotation
\begin{equation}\label{di}D^{\alpha\beta}(\textbf{r},t)=\int dR\sum_{i}\delta(\textbf{r}-\textbf{r}_{i})d_{i}^{\alpha}d_{i}^{\beta}\psi^{*}(R,t)\psi(R,t)\end{equation}
 in the equation (\ref{di eq for pol current selfconsist appr}).

It can be assumed that
\begin{equation}\label{di appr for D funct}D^{\alpha\beta}(\textbf{r},t)\simeq\sigma\frac{P^{\alpha}(\textbf{r},t)P^{\beta}(\textbf{r},t)}{n(\textbf{r},t)}.\end{equation}
where $\sigma$- is the nondimension constant.

Likewise the previous section we can reformulate the
equation (\ref{di eq for pol current selfconsist appr})
for the three-dimension case in terms of the electrical
field (\ref{di field eq}). The
equation (\ref{di eq for pol current selfconsist appr})
in this situation has the form
$$\partial_{t}R^{\alpha\beta}(\textbf{r},t)+\frac{1}{m}\partial^{\gamma}R^{\alpha\beta\gamma}(\textbf{r},t)=\frac{e}{m}P^{\alpha}(\textbf{r},t)E^{\beta}(\textbf{r},t)$$
\begin{equation}\label{di eq for pol current selfconsist appr 3D} +\frac{e}{mc}\varepsilon^{\beta\gamma\delta}R^{\alpha\gamma}(\textbf{r},t)B_{ext}^{\delta}(\textbf{r},t)+\frac{1}{m}D^{\alpha\gamma}(\textbf{r},t)\partial^{\beta}E^{\gamma}(\textbf{r},t).
\end{equation}

If we put a velocity field into the
equation (\ref{di eq for pol current selfconsist appr})
the tensor $R^{\alpha\beta\gamma}(\textbf{r},t)$ transforms into  $$R^{\alpha\beta\gamma}(\textbf{r},t)=r^{\alpha\beta\gamma}(\textbf{r},t)+T^{\alpha\beta\gamma}(\textbf{r},t)$$
$$+mR^{\alpha\beta}(\textbf{r},t)v^{\gamma}(\textbf{r},t)+mR^{\alpha\gamma}(\textbf{r},t)v^{\beta}(\textbf{r},t)$$
\begin{equation}\label{di current of polarization with vel field}-mP^{\alpha}(\textbf{r},t)v^{\beta}(\textbf{r},t)v^{\gamma}(\textbf{r},t).\end{equation}

In this representation the contribution of thermal movement
\begin{equation}\label{di pressure polariz} r^{\alpha\beta\gamma}(\textbf{r},t)=\int dR\sum_{i=1}^{N}\delta(\textbf{r}-\textbf{r}_{i})d_{i}^{\alpha}a^{2}(R,t)m_{i}u^{\beta}_{i}u^{\gamma}_{i} \end{equation}
becomes explicit and the analog for the Bohm quantum potential occurs
 $$T^{\alpha\beta\gamma}(\textbf{r},t)=-\frac{\hbar^{2}}{2m}\int dR\sum_{i=1}^{N}\delta(\textbf{r}-\textbf{r}_{i})\times$$
\begin{equation}\label{di Bom1 polariz} \times d_{i}^{\alpha}a^{2}(R,t)\frac{\partial^{2}\ln a}{\partial x_{\beta i}\partial x_{\gamma i}}
.\end{equation}
Approximate connection of $T^{\alpha\beta\gamma}(\textbf{r},t)$
with the concentration and polarization of particles is
presented in work ~\cite{Andreev arxiv 11 2}.

The major contribution into alterations of polarization
in a system of charged particles is from charge interactions
and from the effect that the external electrical field has on
charges. As a result
$$\partial_{t}R^{\alpha\beta}(\textbf{r},t)
=\frac{e}{m}P^{\alpha}(\textbf{r},t)E_{ext}^{\beta}(\textbf{r},t)$$
\begin{equation}\label{di eq for pol current selfconsist appr charge}+\frac{e^{2}}{m}P^{\alpha}(\textbf{r},t)\partial^{\beta}\int d\textbf{r}'G(\textbf{r},\textbf{r}')n(\textbf{r}',t)\end{equation}
can be derived from the equation (\ref{di eq for pol current selfconsist appr}).

Later in this paper the equation (\ref{di eq for pol current selfconsist appr charge})
is used to analyze the elementary excitations spectrum in a two-dimensional
system of EDM-having charged particles.

Method of QHD allows derive an one-particle nonlinear Schrodinger
equation (NLSE). Corresponding one-particle wave function describes
evolution of many-particle system in 3D physical space.  Most
famous NLSE are the Ginzburg-Landau equation in superconductivity
~\cite{Rosenstein RMP 10} and the Gross-Pitaevskii equation in the
physics of ultracold gases and Bose-Einstein condensate
~\cite{Dalfovo RMP 99}. In this article we consider the derivation
of NLSE. We do not use NLSE for studying of wave dispersion and
wave generation. Because of it we present NLSE in Appendix. In
this paper we receive NLSE for the charged particle having EDM.
Interesting consequence of derivation of NLSE is the obtaining of
London equation. Derivation of the London's equation from QHD also
is presented in appendix

\section{\label{sec:level1} IV. The eigenwaves in a 2D system of charged particles with EDM}

One of the primary goals of this article is to derive dispersion characteristics of eigenwaves in 2D systems of neutral and charged particles that take account of the EDM collective dynamics. In this section we consider 2D systems of charged particles. To do that let's analyze small perturbations of physical variables from the stationary state. The ionic or holes gas is assumed here to localize on a  $x,y$-plane
\begin{equation}\label{di}\begin{array}{ccc}n=n_{0}+\delta n,& v^{\alpha}=0+\delta v^{\alpha},& P^{\alpha}=P_{0}^{\alpha}+\delta P^{\alpha},\end{array}
\end{equation}
$$\begin{array}{ccc}E_{ext}^{\alpha}=E_{0}\delta^{\alpha z},& P_{0}^{\alpha}=\kappa E_{0}^{\alpha}=\kappa E_{0}\delta^{\alpha z},& \end{array}
$$
where $n$ is a surface concentration.

We can mark, in this section and below, all physical quantity is presented in the form of sum of equilibrium part and small perturbations
$$f=f_{0}+\delta f.$$

In this case if we assume that linear excitations $\delta f$ are
proportional to $exp(-\imath\omega t+\imath\textbf{k}\textbf{r})$
a linearized set of equations (\ref{di cont eq}), (\ref{di bal imp
eq}), (\ref{di eq polarization}), (\ref{di eq for pol current
selfconsist appr charge}) gives us the dispersion equation
$$m\omega^{2}=mv_{si}^{2}k^{2}+\frac{\hbar^{2}}{4m}k^{4}$$
\begin{equation}\label{di disp eq}+2\pi e^{2}n_{0}k-2\pi P_{0}^{2}\frac{\beta(k) e^{2}}{m\omega^{2}}k^{4},\end{equation}
where
\begin{equation}\label{di const two dim}\beta(k)=2\pi\int_{\xi}^{\infty}dr\frac{J_{0}(r)}{r^{2}},\end{equation}
here $\xi=r_{0}k$, $r_{0}$ there is an ionic or atomic radius and $k=\sqrt{k_{x}^{2}+k_{y}^{2}}$ is a modulus of the wave vector. As $\lambda_{min}=2\pi/k_{max}>2r_{0}$ then $\xi\subset(0,\pi)$. The function $\beta(\xi)$ is, therefore, positive. Its explicit form is presented at Fig. 1.

Solutions of the equation (\ref{di disp eq}) are presented by two branches of the dispersion characteristic
$$\omega^{2}=\frac{1}{2}\Biggl(v_{si}^{2}k^{2}+\frac{\hbar^{2}}{4m^{2}}k^{4}+\omega_{Li2}^{2}$$
\begin{equation}\label{di disp general for charged part}\pm\sqrt{(v_{si}^{2}k^{2}+\frac{\hbar^{2}}{4m^{2}}k^{4}+\omega_{Li2}^{2})^{2}-8\pi\kappa^{2}E_{0}^{2}\beta\frac{e^{2}}{m^{2}}k^{4}}\Biggr).\end{equation}
where $\omega_{Li2}=2\pi e^{2}n_{0}k/m$ denotes the Langmuir 2D frequency.

Provided that polarization effects are small the square root expression in (\ref{di disp general for
charged part}) may be expanded to yield formulae $\omega(k)$ in the following form
$$\omega^{2}=v_{si}^{2}k^{2}+\frac{\hbar^{2}}{4m^{2}}k^{4}$$
\begin{equation}\label{di disp for charged part Langmuir}+\omega_{Li2}^{2}-\frac{2\pi\kappa^{2}E_{0}^{2}\beta e^{2}k^{4}}{m^{2}v_{si}^{2}k^{2}+\frac{1}{4}\hbar^{2}k^{4}+m^{2}\omega_{Li2}^{2}},\end{equation}
\begin{equation}\label{di disp for charged part dipole}\omega^{2}=\frac{2\pi\kappa^{2}E_{0}^{2}\beta e^{2}k^{4}}{m^{2}v_{si}^{2}k^{2}+\frac{1}{4}\hbar^{2}k^{4}+m^{2}\omega_{Li2}^{2}}.\end{equation}

The relation (\ref{di disp for charged part Langmuir}) expresses the dispersion of 2D Langmuir waves with the account of dipole-dipole interactions. The solution (\ref{di disp for charged part dipole}) expresses the dispersion of waves that emerge as a result of polarization dynamics. Plots at Figs. 2 and 3 represent relationships (\ref{di disp for charged part Langmuir}), (\ref{di disp for charged part dipole}), respectively.

\begin{figure}
\includegraphics[width=8cm,angle=0]{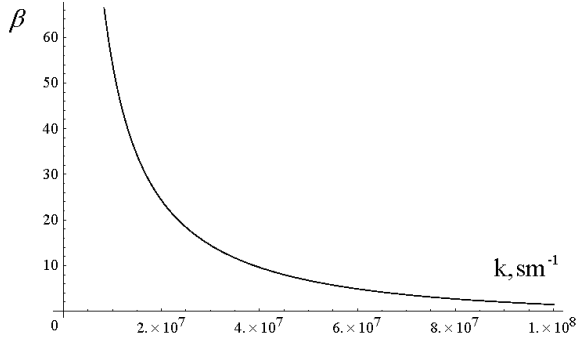}
\caption{\label{fig1PCaNP:epsart} The figure presents the
dependence of the variable $\beta(k)$ on the wave vector $k$.
Ionic or atomic radius $r_{0}$ is assumed to equal 0.1 nm.}
\end{figure}

\begin{figure}
\includegraphics[width=8cm,angle=0]{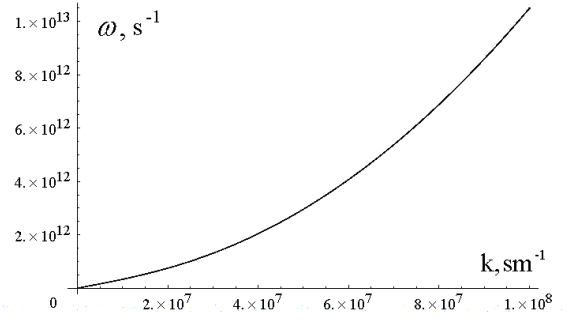}
\caption{\label{fig2PCaNP:epsart} The figure shows the dispersion
characteristic of the Langmuir wave frequency $\omega$ versus the
wave vector $k$, which is described by the equation (\ref{di disp
for charged part Langmuir}). The ionic radius $r_{0}$ is supposed
to be 0.1 nm. Equilibrium polarization has form $P_{0}=\kappa
E_{0}$. Static electric permeability $\kappa$ is defined by the
equation $\kappa=n_{0}p_{0}^{2}/(3k_{B}T)$. $p_{0}$ - is a dipole
moment of an ion or atom, $T$ - temperature of the medium, $k_{B}$
- Boltzmann constant. System parameters are assumed to be as
follows: $n_{0}=10^{8}sm^{-2}$, $p_{0}=3\cdot 10^{-20}C m$,
$T=100K$, $E_{0}=3\cdot 10^{4}V/m$, $m_{i}=10^{-23}g$.}
\end{figure}

\begin{figure}
\includegraphics[width=8cm,angle=0]{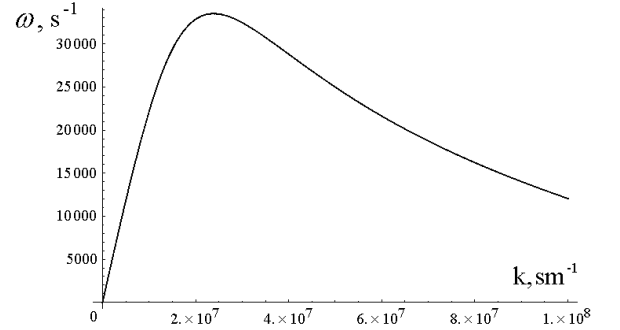}
\caption{\label{fig3PCaNP:epsart} The figure shows
the dispersion characteristic of the quantum ionic polarization wave
frequency $\omega$ versus the wave vector $k$, which is described by the
equation (\ref{di disp for charged part dipole}).
The ionic radius $r_{0}$ is supposed to be 0.1 nm. Physical conditions
of the system are supposed to be the same as described in the legend to Fig. 2.}
\end{figure}

\section{\label{sec:level1} V. Eigenwaves in a system of polarized neutral particles}

In this section we consider 1D, 2D and 3D systems of EDM-having
neutral particles. To do that we use the equation (\ref{di eq for pol current selfconsist
appr}). Since the system of neutral particles resides in a
uniform electromagnetic field, there is only one term on
the right side of the equation (\ref{di eq for pol current selfconsist appr}).
It is also assumed that interactions make the largest contribution into the changes in $R^{\alpha\beta}(\textbf{r},t)$.

If so then the equation (\ref{di eq for pol current selfconsist
appr}) and (\ref{di appr for D funct}) transforms into
$$\partial_{t}R^{\alpha\beta}(\textbf{r},t)=\sigma\frac{P^{\alpha}(\textbf{r},t)P^{\gamma}(\textbf{r},t)}{mn(\textbf{r},t)}\times$$
\begin{equation}\label{di eq for pol current selfconsist appr neutral}\times\partial^{\beta}\int d\textbf{r}'G^{\gamma\delta}(\textbf{r},\textbf{r}')P^{\delta}(\textbf{r}',t).\end{equation}

If taken in a linear approximation, the set of equations (\ref{di cont eq}),
(\ref{di bal imp eq}), (\ref{di eq polarization}), (\ref{di eq for
pol current selfconsist appr neutral}) for neutral particles is closed. This allows the analysis of polarization waves in a system of neutral particles. Equations of continuity (\ref{di cont eq}) and of the momentum balance (\ref{di bal imp eq}) herein describe the dynamics of acoustic wave.
If we derive a solution for eigenwaves in a 2D system the dispersion equation has a form of
\begin{equation}\label{di disp for neutral part}\omega=\sqrt{\sigma\frac{\beta(k) }{mn_{0}}}\mid\kappa\mid E_{0}k^{3/2}.\end{equation}
where $\beta(k)$ is defined by the relation (\ref{di const two dim}). The dispersion dependence (\ref{di disp for neutral part}) is presented on Fig. 4.

In 1D case $\omega(k)$ occurs as
\begin{equation}\label{di disp for neutral part one dim}\omega=\sqrt{\frac{\sigma\beta_{1}(k)}{mn_{0}}}\mid\kappa\mid E_{0}k^{2},\end{equation}
where
\begin{equation}\label{di koef one dim}\beta_{1}(k)=2\int_{\xi}^{\infty}dr\frac{cos(r)}{r^{3}}.\end{equation}
The quantity (\ref{di koef one dim}) and 1D dispersion dependence (\ref{di disp for neutral part one dim}) are presented on Figs. (5) and (6) correspondingly.

The set of equations (\ref{di eq polarization}), (\ref{di eq for pol current selfconsist appr neutral}), (\ref{di appr for D funct}) may be applied to analyze 3D systems, otherwise the electric field $\textbf{E}(\textbf{r},t)$ that dipoles create may be introduced explicitly (\ref{di field eq}), (\ref{di eq polarization}), (\ref{di eq for pol current selfconsist appr 3D}), (\ref{di appr for D funct}).

As this is done the dispersion equation $\omega(k)$ transforms into
\begin{equation}\label{di disp for neutral part three dim}\omega=\sqrt{\frac{4\pi\sigma}{mn_{0}}}P_{0}k_{z}.\end{equation}

Equations (\ref{di disp for neutral part}) and (\ref{di disp for
neutral part one dim}) differ in the power coefficient by the wave
vector $k$ and also with the coefficients $\beta(k)$ and
$\beta_{1}(k)$, which depend on the wave vector, occur in 1D and
2D cases, respectively.

\begin{figure}
\includegraphics[width=8cm,angle=0]{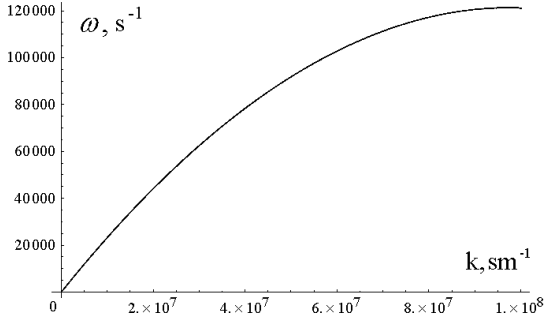}
\caption{\label{fig4PCaNP:epsart} The figure shows the dispersion
characteristic of the 2D quantum atomic polarization wave frequency $\omega(k)$
versus the wave vector $k$, which is described by the equation (\ref{di disp for neutral part}).
The atomic radius is supposed to be 0.1 nm. System parameters are the
same as in Fig. 2.}
\end{figure}

\begin{figure}
\includegraphics[width=8cm,angle=0]{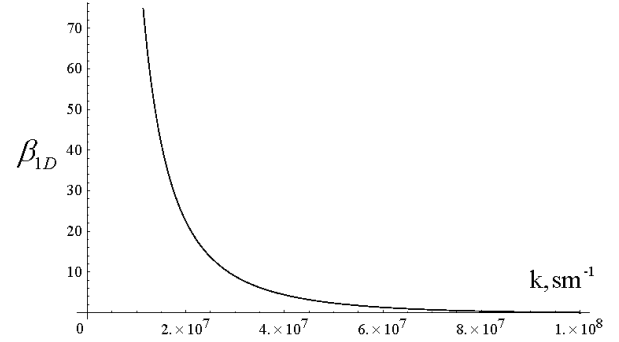}
\caption{\label{fig5PCaNP:epsart} The figure presents of the
variable $\beta_{1}(k)$ on the wave vector $k$. Ionic or atomic radius $r_{0}$ are
assumed to equal 0.1 nm.}
\end{figure}

\begin{figure}
\includegraphics[width=8cm,angle=0]{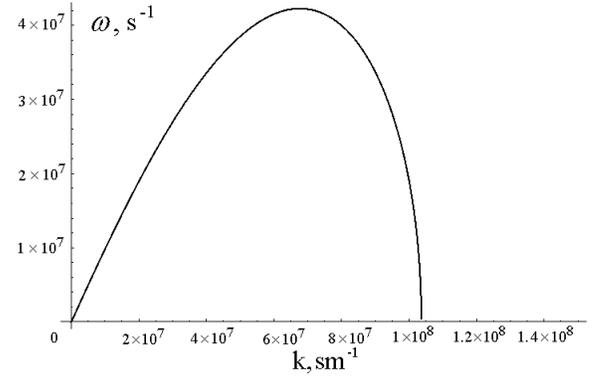}
\caption{\label{fig5PCaNP:epsart} The dependence of frequency
$\omega$ on the wave vector $k$ is displayed for the case of
single dimension polarization mode which dispersion characteristic
is defined by the equation (\ref{di disp for neutral part one
dim}). The atomic radius $r_{0}$ is supposed to be 0.1 nm.
Equilibrium polarization has form $P_{0}=\kappa E_{0}$. Static
electric permeability $\kappa$ is defined by the equation
$\kappa=n_{0}p_{0}^{2}/(3k_{B}T)$. $p_{0}$ - is a dipole moment of
an atom, $T$ - temperature of the medium, $k_{B}$ - Boltzmann
constant. System parameters are assumed to be as follows:
$n_{0}=10^{4}sm^{-1}$, $p_{0}=3\cdot 10^{-20}C m$, $T=100K$,
$E_{0}=3\cdot 10^{4}V/m$, $m=10^{-23}g$.}
\end{figure}

\section{\label{sec:level1} VI. Eigenwaves in a 2D two-sort system of charged particles}

A 2D system comprised by charged particles of two sorts can be
modeled either in a thin film of metal and semiconductors or in a thin, nano-sized
ionic crystal. Only ions and holes can have EDM in the former
case. Both sorts of ions can have EDM in the case of ionic
crystal. In our analysis we deal with a material where one sort of
ions has much greater EDM than the other and investigate
elementary excitations in such system.

The QHD equation set for the system in question is
comprised by continuity equations (\ref{di cont eq})
and momentum balance equations (\ref{di current to velocity})
for both sorts of particles and also by equations (\ref{di eq polarization}), (\ref{di eq for pol current
selfconsist appr}) , which describe the polarization
dynamics of particles those having greater EDM. The
field on the right side of equations (\ref{di current to velocity}) and (\ref{di
eq for pol current selfconsist appr}) comes from
particles of both sorts. Likewise previous sections
the 2D system is assumed here to be located in an
external electric field which is orthogonal to
the plane of particle movement. Solution of the
eigenwave problem in a two-sort system of charged particles leads to the dispersion equation
$$(\omega^{2})^{3}-\Biggl(\omega_{e}^{2}+v_{se}^{2}k^{2}+\frac{\hbar^{2}k^{4}}{4m_{e}^{2}}+\omega_{i}^{2}
+v_{si}^{2}k^{2}+\frac{\hbar^{2}k^{4}}{4m_{i}^{2}}\Biggr)(\omega^{2})^{2}$$
$$+\Biggl((v_{se}^{2}+\frac{\hbar^{2}k^{2}}{4m_{e}^{2}})
(v_{si}^{2}+\frac{\hbar^{2}k^{2}}{4m_{i}^{2}})k^{4}+\omega_{i}^{2}(v_{se}^{2}+\frac{\hbar^{2}k^{2}}{4m_{e}^{2}})k^{2}$$
$$+\omega_{e}^{2}(v_{si}^{2}+\frac{\hbar^{2}k^{2}}{4m_{i}^{2}})k^{2}
+\frac{\beta(k)\omega_{i}^{2}P_{0}^{2}k^{4}}{n_{0}m_{i}}\Biggr)\omega^{2}$$
\begin{equation}\label{di disp eq two sorts}-\frac{\beta(k)
P_{0}^{2}\omega_{i}^{2}}{n_{0}m_{i}}(v_{se}^{2}+\frac{\hbar^{2}k^{2}}{4m_{e}^{2}})k^{6}=0.\end{equation}

If we analyze a film of metal the indices $e$ and $i$ relate to electrons and
ions, respectively. For the case of semiconductors index $i$ describes a holes.
In the case of a nanolayer ionic crystal $i$ relates
to ions of different sorts. It is assumed in both cases that particles
denoted by the indices $e$ and $i$ are those having EDM.

If there is
no polarization then the equation (\ref{di disp eq two sorts}) is a
second-degree equation with respect to $\omega^{2}$. Its solutions have the form
$$\omega^{2}_{0\pm}=\frac{1}{2}\Biggl(\omega_{e}^{2}+\omega_{i}^{2}+v_{qse}^{2}k^{2}
+v_{qsi}^{2}k^{2}$$
\begin{widetext}
\begin{equation}\label{di like known wave sol}\pm\sqrt{(\omega_{e}^{2}+\omega_{i}^{2})^{2}+(v_{qse}^{2}k^{2}
-v_{qsi}^{2}k^{2}
)^{2}+(\omega_{e}^{2}-\omega_{i}^{2})(v_{qse}^{2}k^{2}
-v_{qsi}^{2}k^{2})}\Biggr).\end{equation}
 \end{widetext}

The following denotation is used here:
$$v_{qsa}^{2}=v_{sa}^{2}+\hbar^{2}k^{2}/4m_{a}^{2},$$
where $a$ denotes $e$ for electrons and $i$ for ions.

The equation (\ref{di
like known wave sol}) contains dispersion characteristics for
the Langmuir wave and ion-acoustic wave. In the quantum case
we consider these characteristics take on the form of
\begin{equation}\label{di Lengm
wave}\omega^{2}=\omega^{2}_{e}+v^{2}_{se}k^{2}+\frac{\hbar^{2}}{4m^{2}_{e}}k^{4},\end{equation}
and
\begin{equation}\label{di ion-sound wave}\omega=\frac{kv_{s}
\sqrt{1+\frac{\hbar^{2}k^{2}}{4m_{e}^{2}v_{se}^{2}}}}
{\sqrt{1+a_{e}^{2}k^{2}(1+\frac{\hbar^{2}k^{2}}{4m_{e}^{2}v_{se}^{2}})}}.\end{equation}
Here used designation $a_{e}^{2}=v_{se}^{2}/\omega_{e}^{2}$,
$v_{s}=(m_{e}/m_{i})v_{se}^{2}$.

In approximation:
\begin{equation}\label{di appr for ion-acc wave}v_{si}^{2}+\hbar^{2}k^{2}/4m_{i}^{2}\ll\omega^{2}/k^{2}\ll
v_{se}^{2}+\hbar^{2}k^{2}/4m_{e}^{2}\end{equation}
formula (\ref{di disp pol
wave in who sorts}) take form:
\begin{equation}\label{di ion-sound wave appr}\omega=\frac{kv_{s}}{\sqrt{1+a_{e}^{2}k^{2}}}\Biggl(1+\frac{\hbar^{2}k^{2}}{8m_{e}^{2}v_{se}^{2}}\frac{1}{1+a_{e}^{2}k^{2}}\Biggr).\end{equation}
In this approximation exist simple relation for ion-sound waves
(\ref{di ion-sound wave}).

The solution (\ref{di ion-sound
wave}) differs from the classical case in the
presence of a term proportional to $\hbar^{2}$
which occurs due to Bohm quantum potential (\ref{di Bom2}).
Figure 7 shows dispersion characteristics of ion-acoustic
waves in classical and quantum cases.

If polarization is taken into account in the equation (\ref{di disp eq two sorts})
then new solution occurs. Assuming the contribution of polarization to be small
we may obtain the following dispersion characteristic for this novel solution:
\begin{widetext}
\begin{equation}\label{di disp pol wave in who sorts}\omega^{2}=\frac{1}{n_{0}m_{i}}
\frac{\beta(k)
P_{0}^{2}\omega_{i}^{2}k^{3}(v_{se}^{2}+\frac{\hbar^{2}k^{2}}{4m_{e}^{2}})}
{\omega_{e}^{2}(v_{si}^{2}+\frac{\hbar^{2}k^{2}}{4m_{i}^{2}})+\omega_{i}^{2}(v_{se}^{2}
+\frac{\hbar^{2}k^{2}}{4m_{e}^{2}})+(v_{se}^{2}+\frac{\hbar^{2}k^{2}}{4m_{e}^{2}})(v_{si}^{2}
+\frac{\hbar^{2}k^{2}}{4m_{i}^{2}})k^{2}}.\end{equation}
\end{widetext}

If the approximation (\ref{di appr for ion-acc wave}) is used then the
equation (\ref{di disp pol wave in who sorts}) takes the form
\begin{equation}\label{di disp pol wave in who sorts appr isw}\omega=\sqrt{\frac{\beta(k)}{n_{0}m_{i}}}P_{0}k^{3/2}.\end{equation}
The equation (\ref{di ion-sound wave appr}) for ion-acoustic
waves occurs in the same approximation.

The polarization also causes additional contributions
in solutions (\ref{di like known wave sol}) to occur.
Thus, a generalization of solutions (\ref{di like known wave sol})
that takes account of polarization appears in the following form:
$$\omega_{\pm}^{2}=\omega_{0\pm}^{2}\mp\frac{1}{\omega_{0+}^{2}-\omega_{0-}^{2}}\times$$
\begin{equation}\label{di like known wave sol and pol}\times\frac{\beta(k) P_{0}^{2}k^{3}\omega_{i}^{2}}{n_{0}m_{i}}\Biggl(1-k^{2}\frac{v_{qse}^{2}}{\omega_{0\pm}^{2}}\Biggr).\end{equation}

\begin{figure}
\includegraphics[width=8cm,angle=0]{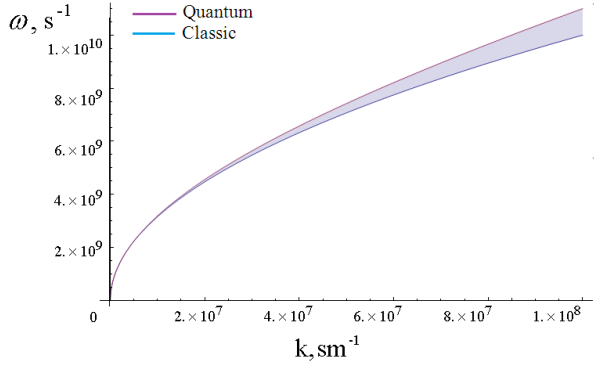}
\caption{\label{fig5PCaNP:epsart} This figure displays the
function $\omega(k)$ for two-dimensional ion-acoustic waves. The
upper curve shows the dispersion characteristic of quantum
ion-acoustic waves which are described by equation (\ref{di
ion-sound wave appr}) while the lower curve shows the dispersion
characteristic of classic ion-acoustic waves which occur from
(\ref{di ion-sound wave appr}) if the term proportional to
$\hbar^{2}$ is neglected. System parameters are assumed to be as
follows: $n_{0}=10^{8}sm^{-2}$, $m=10^{-23}g$.}
\end{figure}

\section{\label{sec:level1} VII. The effect of polarization on the dispersion characteristics of 2D semiconductors}

The simplest model of semiconductors is a system comprised by
particles of three sorts, namely electrons, ions and neutral
atoms. Atoms and ions in the semiconductor may bear significant
EDM. Dynamics of semiconductors is usually dealt with as dynamics
of electrons and holes. Electrons and holes can also form bound
states - the excitons. As excitons have EDM, the dipole-dipole
interaction plays a substantial role in a system of excitons.

The approach we present in this paper allows us to describe the
interaction of excitons via their EDMs and interaction of their
EDM with an external electric field. Since excitons are neutral
EDM-having quasi-particles, the result we obtained in Section VI
may be used to describe elementary excitations in such system. We
can therefore suggest that acoustic-like waves and polarization
waves may be excited in exciton systems and their dispersion is
described by formulaes (\ref{di disp for neutral part}), (\ref{di
disp for neutral part one dim}) and (\ref{di disp for neutral part
three dim}).

In the case electrons and holes move in a semiconductor without
significant formation of bound state excitations in a
semiconductor may be regarded to as excitations in a two-sort
system of charged particles. The term "hole" regards to ions in
the semiconductor, which may have EDM. This means that holes may
be dealt with as EDM-having particles.

Using the model described above we can apply formulae from the
Section VII to describe the dispersion of elementary excitations in
2D electron-hole plasma. Index $i$ in formulas
(\ref{di disp eq two sorts})-(\ref{di like known wave sol and pol}) regards to holes.

\section{\label{sec:level1} VIII. Excitation of polarization waves by beam of neutral polarized particles}

We do not meet experimental works there were obtained the
polarization waves received in our article. On example of 3D
sample we show the way to obtain such waves in experiment. A
possible way of generation of waves of polarization it is the
propagation of beam of electrons or neutral polarized particles
through the sample. Instead of the beam can be considered a
current of particles through the sample, in context of next
section it can be the electrical current. In this and next
sections we show there are instabilities which led to generation
of waves in the sample. The effect of waves generation by beam of
electrons is well in plasma physics ~\cite{Bret PRE 04},
~\cite{Miller BOOK}.

In this section we use equations (\ref{di cont eq}),
(\ref{di bal imp eq short}), (\ref{di eq polarization})
and (\ref{di eq for pol current selfconsist appr 3D})
for each sorts of particles and equation of field
(\ref{di field eq}). In right side of equation
(\ref{di field eq}) the polarization $\textbf{P}$ is a sum
of polarization  of medium $\textbf{P}_{d}$ and polarization of the beam $\textbf{P}_{b}$
$$\textbf{P}=\textbf{P}_{d}+\textbf{P}_{b}.$$

The equilibrium state of system is
characterized by following values of the medium parameters:
$$\begin{array}{ccc}n_{d}=n_{0d}+\delta n_{d},& v^{\alpha}_{d}=0+v^{\alpha}_{d},& \end{array}
$$
\begin{equation}\label{di equlib state}\begin{array}{ccc}& & P^{\alpha}_{d}=P_{0 d}^{\alpha}+\delta P^{\alpha}_{d}, R^{\alpha\beta}_{d}=0+\delta R^{\alpha\beta}_{d}\end{array}
\end{equation}
and values of the beam parameters:
$$\begin{array}{ccc}n_{b}=n_{0b}+\delta n_{b},& v^{\alpha}_{b}=U\delta^{z\alpha}+\delta v^{\alpha}_{b},& \end{array}
$$
\begin{equation}\label{di equlib state beam}\begin{array}{ccc} & P_{b}^{\alpha}=P_{0b}^{\alpha}+\delta P^{\alpha}_{b},& R^{\alpha\beta}_{b}=R^{\alpha\beta}_{0b}+\delta R^{\alpha\beta}_{b}.\end{array}
\end{equation}
The polarization $P_{0}^{\alpha}$ is proportional to external
electric field $E_{0}^{\alpha}$. We consider the case then
$\textbf{E}_{0}=[E_{0}sin\varphi, 0, E_{0}cos\varphi]$. In this
case the tensor $R^{\alpha\beta}_{0b}$ has only two unequal to
zero elements: $R^{zx}_{0b}=R_{0b}sin\varphi$ and
$R^{zz}_{0b}=R_{0b}cos\varphi$. And we consider the small
perturbations of described here equilibrium state.

\begin{equation}\label{di}1-\frac{\omega_{D}^{2}}{\omega^{2}-\frac{\hbar^{2}k^{4}}{4m_{d}^{2}}}-\frac{\omega_{Db}^{2}}{(\omega-k_{z}U)^{2}-\frac{\hbar^{2}k^{4}}{4m_{b}^{2}}}=0,\end{equation}

In the absence of medium we obtain a dispersion dependence of polarized beam modes
\begin{equation}\label{di}\omega=k_{z}U\pm\sqrt{\omega_{Db}^{2}+\frac{\hbar^{2}k^{4}}{4m_{e}^{2}}}.\end{equation}

At the presence of medium, far from resonant condition $k_{z}U\simeq \omega_{Db}$, we obtain
\begin{equation}\label{di beam with medium NOres DIP beam}\omega\simeq k_{z}U\pm\frac{\omega_{Db}}{\sqrt{1-\frac{\omega_{D}^{2}}{(k_{z}U)^{2}}}}=k_{z}U\pm\imath\frac{\omega_{Db}}{\sqrt{\frac{\omega_{D}^{2}}{(k_{z}U)^{2}}-1}}\end{equation}
For the case $\omega_{D}\gg k_{z}U$ from (\ref{di beam with medium NOres DIP beam}) we derive
$$\omega=k_{z}U\biggl(1\pm\imath\frac{\omega_{Db}}{\omega_{D}}\biggr).$$
Further we consider resonant interaction neutral polarized beam with medium. In this situation we present frequency in the form
$$\omega=k_{z}U+\eta.$$
At resonant condition $\omega_{D}\simeq k_{z}U$ and
$\eta\gg\frac{\hbar k^{2}}{m_{b}}$ we have
\begin{equation}\label{di}\eta=\xi\sqrt[3]{\frac{\omega_{Db}^{2}\omega_{D}^{2}}{2\sqrt{\omega_{D}^{2}+\frac{\hbar^{2}k^{4}}{4m_{d}^{2}}}}}.\end{equation}
where $\xi$ presented by formula (\ref{di xi sqrt 3}). In the
limit $\eta\ll\frac{\hbar k^{2}}{m_{b}}$ we obtain a formula which
is analogous to (\ref{di small inkr el}). But in this case we need
to do some changes, i.e. $m_{e}\rightarrow m_{b}$ and
$\omega_{Le}$ to $\omega_{Db}$.

\section{\label{sec:level1} IX. Excitation of polarization waves by electron beam}

In previous section we consider  the effect of generation  of the
polarization waves  in 3D system of neutral particles with the
EDM. Here we analyse an analogous effect. It is a generation of
waves by means of a beam of neutral polarized particles.

Let's analyze the interaction of  a single velocity beam of
electrons that moves along z-axis with infinite three-dimension
medium comprised of neutral particles. Particles of the medium are
in equilibrium state and polarized without any external electric
field. This may occur if the electron beam moves through a crystal
of piezoelectric or through a sample of ferroelectric that has
residual polarization. The result of such analysis would also
represent a physical mechanism of the interaction of an electron
beam with spatially confined or low-dimension polarized systems.
The most significant types of interaction in such systems are the
Coulomb interaction of electrons in the beam, the dipole-dipole
interaction in the medium and the charge-dipole interaction of
dipoles in the medium with the electron beam. Let's assume that
the beam moves alongside the polarization vector and the beam's
velocity doesn't change significantly while it moves through the
sample.

In this section we use equations (\ref{di cont eq}), (\ref{di bal
imp eq short}), (\ref{di eq polarization}) and (\ref{di eq for pol
current selfconsist appr 3D}) for each sorts of particles and
equation of field (\ref{di field eq}).

If so, solution of the QHD equation set leads to the dispersion
equation
\begin{equation}\label{di}1-\frac{\omega_{Le}^{2}}{(\omega-k_{z}U)^{2}-\frac{\hbar^{2}k^{4}}{4m_{e}^{2}}}-\frac{\omega_{D}^{2}}{\omega^{2}-\frac{\hbar^{2}k^{4}}{4m_{d}^{2}}}=0,\end{equation}
where
$$\omega_{D}^{2}=\frac{4\pi\sigma P_{0}^{2}k^{2}}{mn_{0}}.$$

This equation has three solutions. If an account is taken of Bohm
quantum potential and no electron beam is present, solution
\begin{equation}\label{di}\omega^{2}=\omega_{D}^{2}+\frac{\hbar^{2}k^{4}}{4m_{d}^{2}}\end{equation}
occurs for the polarization wave, while two beam modes occur in
the absence of the medium
\begin{equation}\label{di}\omega=k_{z}U\pm\sqrt{\omega_{Le}^{2}+\frac{\hbar^{2}k^{4}}{4m_{e}^{2}}}.\end{equation}

If the contribution from the Bohm quantum potential is neglected
and the limit of a low-density beam is assumed the solution for
beam mode looks like
\begin{equation}\label{di}\omega\simeq k_{z}U.\end{equation}

If we consider the effect of the dipole-comprised medium on the
beam mode the following solution is obtained
\begin{equation}\label{di beam with medium NOres}\omega\simeq k_{z}U\pm\frac{\omega_{Le}}{\sqrt{1-\frac{\omega_{D}^{2}}{(k_{z}U)^{2}}}}=k_{z}U\pm\imath\frac{\omega_{Le}}{\sqrt{\frac{\omega_{D}^{2}}{(k_{z}U)^{2}}-1}}\end{equation}

This solution is complex and leads to the instability if
$\omega_{D}>k_{z}U$ i.e. in the case of long wave approximation.

If $\omega_{D}\gg k_{z}U$ then solution transforms to
\begin{equation}\label{di}\omega=k_{z}U\biggl(1\pm\imath\frac{\omega_{Le}}{\omega_{D}}\biggr).\end{equation}

Solution (\ref{di beam with medium NOres}) is valid for a beam
moving in a 3D system of dipoles if
\begin{equation}\label{di}k_{z}U\neq\omega_{D},\end{equation}
i.e. if there is no resonance between beam mode and polarization
mode.

Let's see how a beam affects the dispersion of polarization wave
in the absence of resonance with beam mode. The dependence of
frequency on the wave vector has the form
\begin{equation}\label{di res cond among beam and dip}k_{z}U=\sqrt{\omega_{D}^{2}+\frac{\hbar^{2}k^{4}}{4m_{d}^{2}}}.\end{equation}

To analyze a resonance interaction that occurs between the
electron beam and a system of electric dipoles, i.e. the situation
of (\ref{di res cond among beam and dip}) we should seek for a
solution that looks like
\begin{equation}\label{di}\omega=k_{z}U+\eta=\sqrt{\omega_{D}^{2}+\frac{\hbar^{2}k^{4}}{4m_{d}^{2}}}+\eta.\end{equation}

In this instance we obtain the following equation of a frequency
shift $\eta$:
\begin{equation}\label{di}\eta^{3}\pm\frac{\hbar k^{2}}{m_{e}}\eta^{2}-\frac{\omega_{Le}^{2}\omega_{D}^{2}}{2\sqrt{\omega_{D}^{2}+\frac{\hbar^{2}k^{4}}{4m_{d}^{2}}}}=0.\end{equation}

If $\eta\gg\frac{\hbar k^{2}}{m_{e}}$ then it has the following
solution:
\begin{equation}\label{di}\eta=\xi\sqrt[3]{\frac{\omega_{Le}^{2}\omega_{D}^{2}}{2\sqrt{\omega_{D}^{2}+\frac{\hbar^{2}k^{4}}{4m_{d}^{2}}}}},\end{equation}
where
\begin{equation}\label{di xi sqrt 3}\xi=\sqrt[3]{1}=\biggl(1,\frac{-1+\imath\sqrt{3}}{2}, \frac{-1-\imath\sqrt{3}}{2}\Biggr).\end{equation}

In the opposite limit of $\eta\ll\frac{\hbar k^{2}}{m_{e}}$ the
frequency shift occurs as
\begin{equation}\label{di small inkr el}\eta=\pm\imath\sqrt{\frac{m_{e}}{\hbar k^{2}}}\frac{\omega_{Le}\omega_{D}}{\sqrt{2}\sqrt[4]{\omega_{D}^{2}+\frac{\hbar^{2}k^{4}}{4m_{d}^{2}}}}.\end{equation}

So we have shown above in this section that the electron beam that
moves in a system of dipoles polarized by external electric field
induces instability and increases amplitude of the polarization
mode. It is assumed that the electron beam moves alongside of the
external electric field and the medium is infinite and
homogeneous. It should be noted that while the electron beam
accelerates infinitely in such conditions we assumed changes of
the beam velocity to be small. This was done because our goal was
to approximate a system of finite length.

\section{\label{sec:level1} X. Conclusions}

In this paper we analyzed wave excitations caused by the EDM
dynamics in systems of charged and neutral particles. Method of QHD was developed for
EDM-having particles. QHD equations are a consequence of MPSE in
which particles' interaction is directly taken into account. In
our work we consider the Coulomb, charge-dipole and dipole-dipole
interactions.  The system of QHD
equations we constructed is comprised by equations of continuity,
of the momentum balance, of the polarization evolution
 and of the polarization current. In our studies of
wave processed we used a self-consistent field approximation of
the QHD equations.

Using QHD equations we analyzed elementary excitations in various
physical systems in a linear approximation. Waves in a 2D gas of
EDM-having charged particles and in a gas of neutral EDM-having
particles in various physical dimensions were considered. A two-sort 2D system of charged
particles where particles of one sort are assumed to have EDM was
also analyzed.

Dispersion branches of a novel type that occurs due to
polarization dynamics were discovered in those physical systems.
Furthermore, the effect of polarization on the dispersion
characteristics of already known wave modes was studied. The
contribution of Bohm quantum potential which is of purely quantum
origin was taken into account in the calculations. Waves of the
electric polarization we discovered possess the following feature:
their frequencies $\omega$ tends to zero provided that
$k\rightarrow 0$.

The polarization modes that have been derived in our work should
be taken into account in calculations of the thermal capacity and
non-linear dynamics of acoustic and ion-acoustic waves.
Polarization modes may also contribute together with phonons to
the process of formation of Cooper pairs in superconductors. These
polarization modes may be applied in the construction of devices
that implement information transfer processes. Excitements in
polarization may be used as an alternative to spin waves or
together with them. It also should be taken
into account that transfer of polarization disturbances plays a
major role in the information transfer in biological systems. Such
processes do not require particles of the medium to possess EDM as
the dynamics of a system of charged particles leads to collective
polarization $P^{\alpha}(\textbf{r},t)$.

The effect of polarization dynamics on the dispersion
characteristics of exciton systems and of electron-hole plasma is
discussed in section VIII. The polarization mode occurred in those
systems too and its dispersion has been analyzed. The existence of
the polarization mode may affect characteristics of spintronic
and nano-electronic devices.

We show the possibility of the process of waves generation in the system of
polarized particles by means of the monoenergetic beam of neutral polarized
particles and the monoenergetic beam of electrons.

Hence, in this work we present the advancement of the QHD approach
to systems of polarized particles and use it to show the
occurrence of polarization waves in various physical systems
comprised by charged and neutral particles.

\section{\label{sec:level1} Appendix}

\subsection{Derivation of the non-linear Schr\"{o}dinger
equation}
In this section we discuss the derivation of NLSE for systems of
charged particles. The derivation is generally similar to the
derivation of the GP equation for systems of neutral bosons in the
state of BEC ~\cite{Andreev PRA08}. The NLSE comes about from the
continuity equation and the Cauchy integral of the momentum
balance equation. The Cauchy integral exists provided that the
velocity field has the form
\begin{equation}\label{di edde free cond}v^{\alpha}(\textbf{r},t)=\frac{1}{m}\partial^{\alpha}\theta(\textbf{r},t)-\frac{e}{mc}A^{\alpha}(\textbf{r},t)\end{equation}
where $\theta(\textbf{r},t)$ is a velocity field potential.

Starting from QHD equations we can derive an equation for
evolution of the model function defined in terms of hydrodynamic
variables. Thus a macroscopic single-particle wave function may be
defined as
\begin{equation}\label{di def psi func} \Phi(\textbf{r},t)=\sqrt{n(\textbf{r},t)}\exp(\frac{\imath}{\hbar}m\theta(\textbf{r},t)).\end{equation}

If we differentiate this function with respect to time and apply
QHD equations then given the absence of EDM the following equation
is obtained:
$$\imath\hbar\partial_{t}\Phi(\textbf{r},t)=\Biggl(-\frac{\hbar^{2}}{2m}\Biggl(\nabla-\frac{\imath e}{\hbar c}\textbf{A}_{ext}(\textbf{r},t)\Biggr)^{2}+e\varphi(\textbf{r},t)$$
\begin{equation}\label{di nlse int}+e^{2}\int d\textbf{r}'\frac{\mid\Phi(\textbf{r}',t)\mid^{2}}{\mid\textbf{r}-\textbf{r}'\mid}+\Gamma(\textbf{r},t)\Biggr)\Phi(\textbf{r},t),\end{equation}
where
\begin{equation}\label{di pressure term in kin appr} \Gamma(\textbf{r},t)=\int_{\textbf{r}_{0}}^{\textbf{r}} d\textbf{r}'\frac{\nabla p(\textbf{r}',t)}{n(\textbf{r}',t)}.\end{equation}

The equation obtained (\ref{di nlse int}) has the form of NLSE.

NLSE (\ref{di nlse int}) describes collective characteristics in a
system of many charged particles. This follows from the derivation
of NLSE and from the definition of a many-particle wave function
$\Phi(\textbf{r},t)$ (\ref{di def psi func}). A nonlinear term
that is proportional to $e^{2}$ depicts the Coulomb interaction.
Function $\Gamma(\textbf{r},t)$ (\ref{di pressure term in kin
appr}) is the contribution of the kinetic pressure and does not
contain any interaction.

Now let's write down the NLSE that arise from QHD equations
(\ref{di cont eq}) and (\ref{di bal imp eq}) taking into account
the EDM of particles:
$$\imath\hbar\partial_{t}\Phi(\textbf{r},t)=\Biggl(-\frac{\hbar^{2}}{2m}\Biggl(\nabla-\frac{\imath e}{\hbar c}\textbf{A}_{ext}(\textbf{r},t)\Biggr)^{2}$$
$$+e\varphi_{ext}(\textbf{r},t)+\Gamma(\textbf{r},t)+C(\textbf{r},t)$$
\begin{equation}\label{di nlse int polariz}+e\partial^{\beta}\int d\textbf{r}'\frac{P^{\beta}(\textbf{r}',t)}{\mid\textbf{r}-\textbf{r}'\mid}\Biggr)\Phi(\textbf{r},t),\end{equation}
here we use designation $\Gamma(\textbf{r},t)$ described by
formula (\ref{di pressure term in kin appr}). Also in (\ref{di
nlse int polariz}) exist action of electric field on polarization
of particles. This effect is presented by function
$C(\textbf{r},t)$, which determined by following formula
$$C(\textbf{r},t)=\int_{\textbf{r}_{0}}^{\textbf{r}}d\textbf{r}'\frac{1}{n(\textbf{r}',t)}$$
$$\times\Biggl(eP^{\beta}(\textbf{r}',t)\nabla '\partial'^{\beta} \int d\textbf{r}'' G(\textbf{r}',\textbf{r}'')n(\textbf{r}'',t)$$
\begin{equation}\label{di nlse int polariz
detail}+P^{\beta}(\textbf{r}',t)\nabla'\int
d\textbf{r}''G^{\beta\gamma}(\textbf{r}',\textbf{r}'')P^{\gamma}(\textbf{r}'',t)
\Biggr).\end{equation}

If we deal with a 3D system of particles we can introduce a
self-consistent electric field into NLSE as that was done in
equations (\ref{di cont eq}), (\ref{di bal imp eq short}) and
(\ref{di field eq}). In such case equations (\ref{di nlse int
polariz}) and (\ref{di nlse int polariz detail}) transform into
\begin{equation}\label{di nlse int polariz short}\imath\hbar\partial_{t}\Phi(\textbf{r},t)=\Biggl(-\frac{\hbar^{2}}{2m}\Biggl(\nabla-\frac{\imath e}{\hbar c}\textbf{A}_{ext}(\textbf{r},t)\Biggr)^{2}\end{equation}
$$+e\varphi(\textbf{r},t)+\Gamma(\textbf{r},t)+C(\textbf{r},t)\Biggr)\Phi(\textbf{r},t).$$

In this case $C(\textbf{r},t)$ has form
\begin{equation}\label{di nlse int polariz
detail short}
C(\textbf{r},t)=\int_{\textbf{r}_{0}}^{\textbf{r}}d\textbf{r}'\frac{1}{n(\textbf{r}',t)}P^{\beta}(\textbf{r}',t)\nabla
E^{\beta}(\textbf{r}',t),\end{equation}

$\textbf{E}=-\nabla\varphi$ in equations (\ref{di nlse int polariz
short}), (\ref{di nlse int polariz detail short}) and this field
satisfies the equation (\ref{di field eq}).

Given the additivity of electromagnetic field the equations
(\ref{di nlse int polariz short}), (\ref{di nlse int polariz
detail short}) can be generalized and $\textbf{A}_{ext}$
substituted with a vector potential $\textbf{A}$  that is created
by both external and internal sources. QHD equations (\ref{di bal
imp eq}), (\ref{di bal imp eq short}), (\ref{di eq for pol current
selfconsist appr}) and (\ref{di eq for pol current selfconsist
appr 3D}) can be generalized in the same way.

\subsection{Some words about London equation}

In many books the London equation is derived from equation which
describe the dynamics of one particle in external field. But, the
London equation describe the properties of many-particle system,
i.e. system of electrons in superconductors. In this section we
present a new way of derivation of the London equation from QHD
equations, which describe collective dynamics of particles. Also,
we obtain generalization of the London equation

Given the absence of a magnetic field the condition (\ref{di edde
free cond}) becomes a condition of vortex-free flow. Assuming the
approximation of a homogeneous medium $n=n_{0}=const$ an equation
\begin{equation}\label{di eddy free conseq} rot(n_{0}\textbf{v}(\textbf{r},t))=-\frac{en_{o}}{mc}rot(\textbf{A}(\textbf{r},t))\end{equation}
 can be derived from (\ref{di edde free cond}).

If we apply equations $n_{0}\textbf{v}=\textbf{j}$,
$rot\textbf{A}=\textbf{B}$, $rot\textbf{B}=4\pi e/c \textbf{j}$,
$div\textbf{B}=0$ a well-known London equation
\begin{equation}\label{di Londons eq cons}\triangle\textbf{B}(\textbf{r},t)=\frac{4\pi e^{2}n_{0}}{mc^{2}}\textbf{B}(\textbf{r},t)\end{equation}
 can be obtained
where
$$\begin{array}{ccc} \frac{4\pi e^{2}n_{0}}{mc^{2}}=\frac{\omega_{Le}^{2}}{c^{2}}=\frac{1}{\lambda_{L}^{2}},& \lambda_{L}=\frac{c}{\omega_{Le}},& \end{array}$$
$\omega_{Le}$ is the Langmuire frequency and $\lambda_{L}$ is the
Londons' constant ~\cite{Kittel Introduction 1999,Ashcroft and
Mermin}.

In the case of an inhomogeneous medium the following relationship
appears from the equation (\ref{di edde free cond})
 $$rot\textbf{j}(\textbf{r},t)=-\frac{e}{mc}rot(n(\textbf{r},t)\textbf{A}(\textbf{r},t))$$
\begin{equation}\label{di eddy free conseq gen}=-\frac{e}{mc}n(\textbf{r},t)\textbf{B}(\textbf{r},t)-\frac{e}{mc}[\nabla n(\textbf{r},t),\textbf{A}(\textbf{r},t)].\end{equation}

This gives us a generalization for the equation (\ref{di Londons
eq cons}):
\begin{equation}\label{di Londons eq cons
gen}\triangle\textbf{B}(\textbf{r},t)=\frac{4\pi
e^{2}n(\textbf{r},t)}{mc^{2}}\textbf{B}(\textbf{r},t)+\frac{e}{mc}[\nabla
n(\textbf{r},t),\textbf{A}(\textbf{r},t)].\end{equation}


\end{document}